\documentclass[12pt,a4paper]{article}
\usepackage{ifthen,placeins} % for conditional statements
\newboolean{pdflatex}
\setboolean{pdflatex}{true} % False for eps figures 

\usepackage{multirow} % table multirows

\newboolean{articletitles}
\setboolean{articletitles}{true} % False removes titles in references

\newboolean{uprightparticles}
\setboolean{uprightparticles}{false} %True for upright particle symbols

\newboolean{inbibliography}
\setboolean{inbibliography}{false} %True once you enter the bibliography

\usepackage[top=1in, bottom=1.25in, left=1in, right=1in]{geometry}

\usepackage{comment}
\usepackage{microtype}
\usepackage{lineno}  % for line numbering during review
\usepackage{xspace} % To avoid problems with missing or double spaces after
\usepackage{caption} %these three command get the figure and table captions automatically small

\usepackage{graphicx}  % to include figures (can also use other packages)
\usepackage{color}
\usepackage{colortbl}
\graphicspath{{./figs/}} % Make Latex search fig subdir for figures

\usepackage{amsmath} % Adds a large collection of math symbols
\usepackage{amssymb}
\usepackage{amsfonts}
\usepackage{upgreek} % Adds in support for greek letters in roman typeset

\newcommand*\patchAmsMathEnvironmentForLineno[1]{%
\expandafter\let\csname old#1\expandafter\endcsname\csname #1\endcsname
\expandafter\let\csname oldend#1\expandafter\endcsname\csname
end#1\endcsname
 \renewenvironment{#1}%
   {\linenomath\csname old#1\endcsname}%
   {\csname oldend#1\endcsname\endlinenomath}%
}
\newcommand*\patchBothAmsMathEnvironmentsForLineno[1]{%
  \patchAmsMathEnvironmentForLineno{#1}%
  \patchAmsMathEnvironmentForLineno{#1*}%
}
\AtBeginDocument{%
\patchBothAmsMathEnvironmentsForLineno{equation}%
\patchBothAmsMathEnvironmentsForLineno{align}%
\patchBothAmsMathEnvironmentsForLineno{flalign}%
\patchBothAmsMathEnvironmentsForLineno{alignat}%
\patchBothAmsMathEnvironmentsForLineno{gather}%
\patchBothAmsMathEnvironmentsForLineno{multline}%
\patchBothAmsMathEnvironmentsForLineno{eqnarray}%
}

\usepackage{hyperref}    % Hyperlinks in references
\usepackage[all]{hypcap} % Internal hyperlinks to floats.

\def\lhcb {\mbox{LHCb}\xspace}

\def\MagUp {\mbox{\em Mag\kern -0.05em Up}\xspace}

\ifthenelse{\boolean{uprightparticles}}%
{

 \def\Peta        {\ensuremath{\upeta}\xspace}

 \def\Pmu         {\ensuremath{\upmu}\xspace}                 
 \def\Pnu         {\ensuremath{\upnu}\xspace}

 \def\Ptau        {\ensuremath{\uptau}\xspace}                 
                  
 \def\Pphi        {\ensuremath{\upphi}\xspace}

 \def\PDelta      {\ensuremath{\Delta}\xspace}                 
 \def\PXi      {\ensuremath{\Xi}\xspace}                 
 \def\PLambda      {\ensuremath{\Lambda}\xspace}                 
 \def\PSigma      {\ensuremath{\Sigma}\xspace}                 
 \def\POmega      {\ensuremath{\Omega}\xspace}                 
 \def\PUpsilon      {\ensuremath{\Upsilon}\xspace}                 
 
 \def\PB      {\ensuremath{\mathrm{B}}\xspace}                 
                  
 \def\PD      {\ensuremath{\mathrm{D}}\xspace}

 \def\PK      {\ensuremath{\mathrm{K}}\xspace}

 \def\PW      {\ensuremath{\mathrm{W}}\xspace}

 \def\PZ      {\ensuremath{\mathrm{Z}}\xspace}                 
                  
 \def\Pb      {\ensuremath{\mathrm{b}}\xspace}                 
 \def\Pc      {\ensuremath{\mathrm{c}}\xspace}                 
                  
 \def\Pe      {\ensuremath{\mathrm{e}}\xspace}

 \def\Pi      {\ensuremath{\mathrm{i}}\xspace}

 \def\Pt      {\ensuremath{\mathrm{t}}\xspace}

}
{

 \def\Peta        {\ensuremath{\eta}\xspace}

 \def\Pmu         {\ensuremath{\mu}\xspace}                 
 \def\Pnu         {\ensuremath{\nu}\xspace}

 \def\Ptau        {\ensuremath{\tau}\xspace}                 
                  
 \def\Pphi        {\ensuremath{\phi}\xspace}

 \mathchardef\PDelta="7101
 \mathchardef\PXi="7104
 \mathchardef\PLambda="7103
 \mathchardef\PSigma="7106
 \mathchardef\POmega="710A
 \mathchardef\PUpsilon="7107
                  
 \def\PB      {\ensuremath{B}\xspace}                 
                  
 \def\PD      {\ensuremath{D}\xspace}

 \def\PK      {\ensuremath{K}\xspace}

 \def\PW      {\ensuremath{W}\xspace}

 \def\PZ      {\ensuremath{Z}\xspace}                 
                  
 \def\Pb      {\ensuremath{b}\xspace}                 
 \def\Pc      {\ensuremath{c}\xspace}                 
                  
 \def\Pe      {\ensuremath{e}\xspace}

 \def\Pi      {\ensuremath{i}\xspace}

 \def\Pt      {\ensuremath{t}\xspace}

}

\makeatletter
\ifcase \@ptsize \relax% 10pt
  \newcommand{\miniscule}{\@setfontsize\miniscule{4}{5}}% \tiny: 5/6
\or% 11pt
  \newcommand{\miniscule}{\@setfontsize\miniscule{5}{6}}% \tiny: 6/7
\or% 12pt
  \newcommand{\miniscule}{\@setfontsize\miniscule{5}{6}}% \tiny: 6/7
\fi
\makeatother

\DeclareRobustCommand{\optbar}[1]{\shortstack{{\miniscule (\rule[.5ex]{1.25em}{.18mm})}
  \\ [-.7ex] $#1$}}

\def\en         {{\ensuremath{\Pe^-}}\xspace}   % electron negative (\em is taken)
\def\ep         {{\ensuremath{\Pe^+}}\xspace}
\def\epm        {{\ensuremath{\Pe^\pm}}\xspace}

\def\mup        {{\ensuremath{\Pmu^+}}\xspace}
\def\mun        {{\ensuremath{\Pmu^-}}\xspace} % muon negative (\mum is taken)

\def\mupm        {{\ensuremath{\Pmu^\pm}}\xspace}

\def\lepton     {{\ensuremath{\ell}}\xspace}

\def\ellell     {\ensuremath{\ell^+ \ell^-}\xspace}

\def\neu        {{\ensuremath{\Pnu}}\xspace}

\def\neut       {{\ensuremath{\neu_\tau}}\xspace}

\def\W      {{\ensuremath{\PW}}\xspace}
\def\Wp     {{\ensuremath{\PW^+}}\xspace}
\def\Wm     {{\ensuremath{\PW^-}}\xspace}

\def\Z      {{\ensuremath{\PZ}}\xspace}

\def\cquark    {{\ensuremath{\Pc}}\xspace}
\def\cquarkbar {{\ensuremath{\overline \cquark}}\xspace}
\def\ccbar     {{\ensuremath{\cquark\cquarkbar}}\xspace}
\def\bquark    {{\ensuremath{\Pb}}\xspace}
\def\bquarkbar {{\ensuremath{\overline \bquark}}\xspace}
\def\bbbar     {{\ensuremath{\bquark\bquarkbar}}\xspace}
\def\tquark    {{\ensuremath{\Pt}}\xspace}
\def\tquarkbar {{\ensuremath{\overline \tquark}}\xspace}
\def\ttbar     {{\ensuremath{\tquark\tquarkbar}}\xspace}

\def\wz {{\ensuremath{\W\!\Z}}\xspace}
\def\zz {{\ensuremath{\Z\!\Z}}\xspace}
\def\wj {{\ensuremath{\W\!+\!\rm{jets}}}\xspace}
\def\zj {{\ensuremath{\Z\!+\!\rm{jets}}}\xspace}
\def\zsj {{\ensuremath{\Z\!+\!\rm{jet}}}\xspace}

\def\wpb {{\ensuremath{\W\!+\!\bbbar}}\xspace}
\def\wppb {{\ensuremath{\Wp\!+\!\bbbar}}\xspace}
\def\wmpb {{\ensuremath{\Wm\!+\!\bbbar}}\xspace}
\def\zpb {{\ensuremath{\Z\!+\!\bbbar}}\xspace}
\def\wpc {{\ensuremath{\W\!+\!\ccbar}}\xspace}
\def\wppc {{\ensuremath{\Wp\!+\!\ccbar}}\xspace}
\def\wmpc {{\ensuremath{\Wm\!+\!\ccbar}}\xspace}

\def\zpc {{\ensuremath{\Z\!+\!\ccbar}}\xspace}

\def\antikt     {\ensuremath{\text{anti-}k_{\mathrm{T}}}\xspace}
\def\ja {{\ensuremath{\mathrm{j}_1}}\xspace}
\def\jb {{\ensuremath{\mathrm{j}_2}}\xspace}

\def\ipl {\ensuremath{\rm IP(\lepton)}\xspace}
\def\mcfm       {\mbox{\textsc{MCFM}}\xspace}
\def\ugb       {\mbox{\textsc{uGB}}\xspace}
\def\fastjet    {\mbox{\textsc{FastJet}}\xspace}
\def\bdtbcl {\ensuremath{\mathrm{BDT}({bc|udsg})}\xspace}
\def\bdtbc  {\ensuremath{\mathrm{BDT}({b|c})}\xspace}
\def\alphas  {\ensuremath{\alpha_{\rm s}}\xspace}
\newcommand{\deltapdf}{{\ensuremath{\delta_{\rm PDF}}}\xspace}
\newcommand{\deltaalphas}{{\ensuremath{\delta_{\alpha s}}}\xspace}

\newcommand{\deltapdfsq}{{\ensuremath{\delta^{2}_{\rm PDF}}}\xspace}
\newcommand{\deltaalphassq}{{\ensuremath{\delta^{2}_{\alpha s}}}\xspace}

  \def\Kbar    {{\kern 0.2em\overline{\kern -0.2em \PK}{}}\xspace}

\def\KorKbar    {\kern 0.18em\optbar{\kern -0.18em K}{}\xspace}

  \def\Dbar    {{\kern 0.2em\overline{\kern -0.2em \PD}{}}\xspace}

\def\DorDbar    {\kern 0.18em\optbar{\kern -0.18em D}{}\xspace}

\def\Bbar    {{\ensuremath{\kern 0.18em\overline{\kern -0.18em \PB}{}}}\xspace}

\def\BorBbar    {\kern 0.18em\optbar{\kern -0.18em B}{}\xspace}

  \def\Y#1S{\ensuremath{\PUpsilon{(#1S)}}\xspace}% no space before {...}!

\def\Lbar        {{\ensuremath{\kern 0.1em\overline{\kern -0.1em\PLambda}}}\xspace}
\def\LorLbar    {\kern 0.18em\optbar{\kern -0.18em \PLambda}{}\xspace}

\def\to                 {\ensuremath{\rightarrow}\xspace}

\def\AT#1     {\ensuremath{A_{\mathrm{T}}^{#1}}\xspace}           % 2

\def\C#1      {\ensuremath{\mathcal{C}_{#1}}\xspace}                       % 9
\def\Cp#1     {\ensuremath{\mathcal{C}_{#1}^{'}}\xspace}                    % 7
\def\Ceff#1   {\ensuremath{\mathcal{C}_{#1}^{\mathrm{(eff)}}}\xspace}        % 9  
\def\Cpeff#1  {\ensuremath{\mathcal{C}_{#1}^{'\mathrm{(eff)}}}\xspace}       % 7
\def\Ope#1    {\ensuremath{\mathcal{O}_{#1}}\xspace}                       % 2
\def\Opep#1   {\ensuremath{\mathcal{O}_{#1}^{'}}\xspace}                    % 7

\newcommand{\tev}{\ifthenelse{\boolean{inbibliography}}{\ensuremath{~T\kern -0.05em eV}\xspace}{\ensuremath{\mathrm{\,Te\kern -0.1em V}}}\xspace}
\newcommand{\gev}{\ensuremath{\mathrm{\,Ge\kern -0.1em V}}\xspace}
\newcommand{\mev}{\ensuremath{\mathrm{\,Me\kern -0.1em V}}\xspace}
\newcommand{\kev}{\ensuremath{\mathrm{\,ke\kern -0.1em V}}\xspace}
\newcommand{\ev}{\ensuremath{\mathrm{\,e\kern -0.1em V}}\xspace}
\newcommand{\gevc}{\ensuremath{{\mathrm{\,Ge\kern -0.1em V\!/}c}}\xspace}
\newcommand{\mevc}{\ensuremath{{\mathrm{\,Me\kern -0.1em V\!/}c}}\xspace}
\newcommand{\gevcc}{\ensuremath{{\mathrm{\,Ge\kern -0.1em V\!/}c^2}}\xspace}
\newcommand{\gevgevcccc}{\ensuremath{{\mathrm{\,Ge\kern -0.1em V^2\!/}c^4}}\xspace}
\newcommand{\mevcc}{\ensuremath{{\mathrm{\,Me\kern -0.1em V\!/}c^2}}\xspace}

\def\mm   {\ensuremath{\rm \,mm}\xspace}

\def\mum  {\ensuremath{{\,\upmu\rm m}}\xspace}

\def\pb {\ensuremath{\rm \,pb}\xspace}

\def\invfb   {\ensuremath{\mbox{\,fb}^{-1}}\xspace}

\def\gsim{{~\raise.15em\hbox{$>$}\kern-.85em
          \lower.35em\hbox{$\sim$}~}\xspace}
\def\lsim{{~\raise.15em\hbox{$<$}\kern-.85em
          \lower.35em\hbox{$\sim$}~}\xspace}

\def\PDF {PDF\xspace}

\def\sqs   {\ensuremath{\protect\sqrt{s}}\xspace}

\def\mjj        {\mbox{$m_{\rm jj}$}\xspace}
\def\ptjj       {\mbox{$p_{\rm T}({\rm jj})$}\xspace}
\def\ptot       {\mbox{$p$}\xspace}
\def\pt         {\mbox{$p_{\rm T}$}\xspace}
\def\ptl        {\mbox{$p_{\rm T}$(\lepton)}\xspace}
\def\ptlj        {\mbox{$p_{\rm T}({{\rm j}_\lepton})$}\xspace}
\def\lj        {\mbox{${{\rm j}_\lepton}$}\xspace}
\def\ptnu       {\mbox{$p^{\rm miss}_{\rm T}$}\xspace}
\def\ptj        {\mbox{$p_{\rm T}({\rm j})$}\xspace}
\def\etaj       {\mbox{$\eta({\rm j})$}\xspace}

\def\evtgen     {\mbox{\textsc{EvtGen}}\xspace}

\def\geant      {\mbox{\textsc{Geant4}}\xspace}

\def\photos     {\mbox{\textsc{Photos}}\xspace}

\def\pythia     {\mbox{\textsc{Pythia}}\xspace}
\def\alpgen     {\mbox{\textsc{Alpgen}}\xspace}

\def\tell1  {TELL1\xspace}
\def\ukl1   {UKL1\xspace}

\newcommand{\ie}{\mbox{\itshape i.e.}\xspace}

\usepackage{cite} % Allows for ranges in citations
\usepackage{mciteplus}

\usepackage{longtable} % only for template; not usually to be used in PAPERs
\usepackage{booktabs}

\begin{document}

%%%%%%%%%%%%%%%%%%%%%%%%%
%%%%% Title     %%%%%%%%%
%%%%%%%%%%%%%%%%%%%%%%%%%
\renewcommand{\thefootnote}{\fnsymbol{footnote}}
\setcounter{footnote}{1}

% %%%%%%% CHOOSE TITLE PAGE--------
% $Id: title-LHCb-PAPER.tex 85745 2016-01-05 10:00:22Z lafferty $
% ===============================================================================
% Purpose: LHCb-PAPER journal paper title page template
% Author: 
% Created on: 2010-09-25
% ===============================================================================

%%%%%%%%%%%%%%%%%%%%%%%%%
%%%%%  TITLE PAGE  %%%%%%
%%%%%%%%%%%%%%%%%%%%%%%%%
\begin{titlepage}
\pagenumbering{roman}

% Header ---------------------------------------------------
\vspace*{-1.5cm}
\centerline{\large EUROPEAN ORGANIZATION FOR NUCLEAR RESEARCH (CERN)}
\vspace*{1.5cm}
\noindent
\begin{tabular*}{\linewidth}{lc@{\extracolsep{\fill}}r@{\extracolsep{0pt}}}
\ifthenelse{\boolean{pdflatex}}% Logo format choice
{\vspace*{-2.7cm}\mbox{\!\!\!\includegraphics[width=.14\textwidth]{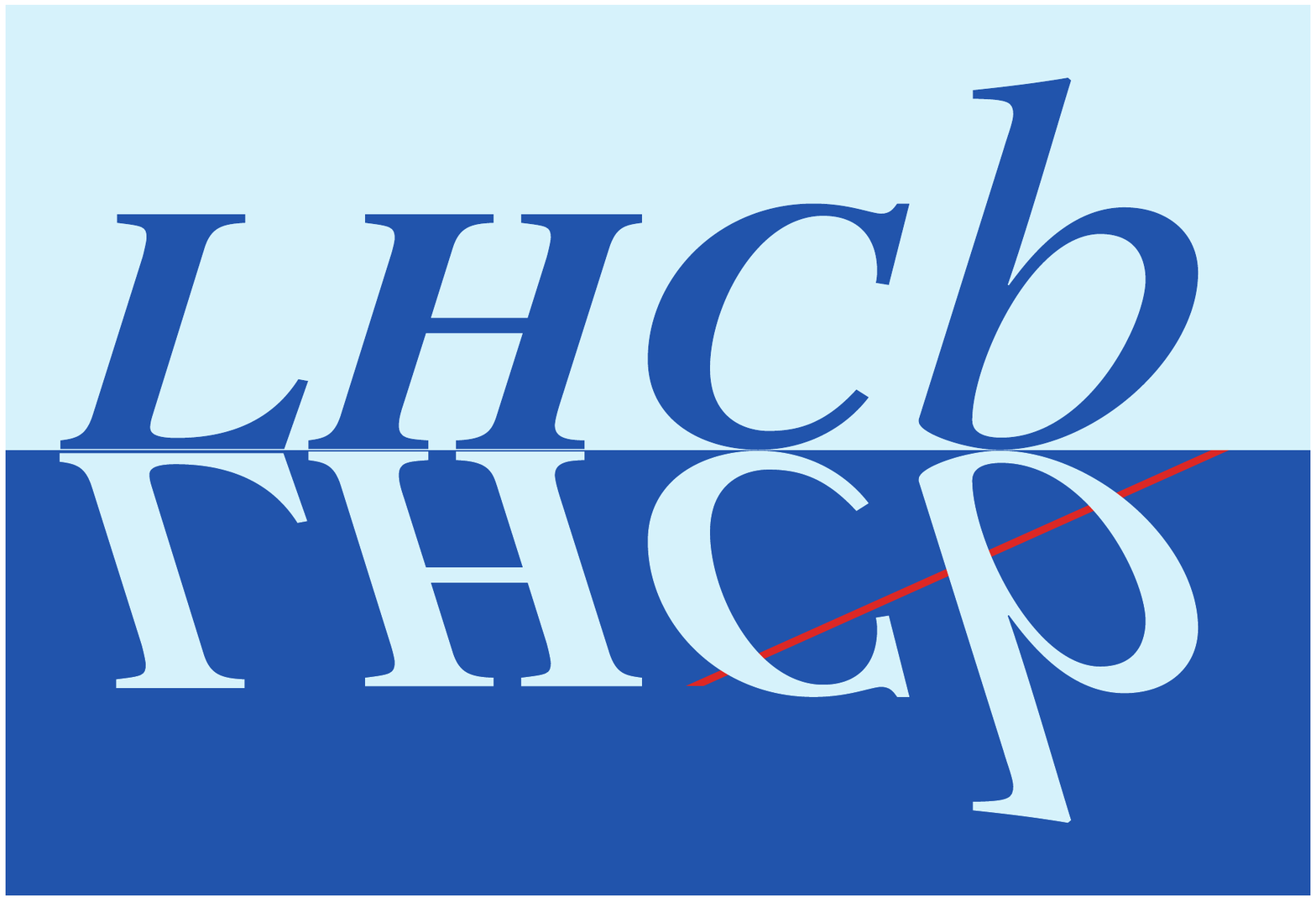}} & &}%
{\vspace*{-1.2cm}\mbox{\!\!\!\includegraphics[width=.12\textwidth]{lhcb-logo.eps}} & &}%
\\
 & & CERN-EP-2016-232 \\  % ID 
 & & LHCb-PAPER-2016-038\\  % ID 
 & & October 25, 2016 \\ % Date - Can also hardwire e.g.: 23 March 2010
 & & \\
% not in paper \hline
\end{tabular*}

\vspace*{4.0cm}

% Title --------------------------------------------------
{\normalfont\bfseries\boldmath\huge
\begin{center}
Measurement of forward \ttbar, \wpb and \wpc production in $pp$ collisions at $\sqrt{s}=8$ TeV
\end{center}
}

\vspace*{2.0cm}

% Authors -------------------------------------------------
\begin{center}
%In the footnote, replace 'paper' by 'letter' in case of submission to PRL or PLB 
The LHCb collaboration\footnote{Authors are listed at the end of this paper.}
\end{center}

\vspace{\fill}

% Abstract -----------------------------------------------
\begin{abstract}
  \noindent
The production of \ttbar, \wpb and \wpc is studied in the forward region of proton-proton collisions collected at a centre-of-mass energy of 8 TeV by the LHCb experiment, corresponding to an integrated luminosity of 1.98 $\pm$ 0.02 \invfb.  The $W$ bosons are reconstructed in the decays $\PW\to\lepton\nu$, where \lepton denotes muon or electron, while the $b$ and $c$ quarks are reconstructed as jets.
All measured cross-sections are in agreement with next-to-leading-order Standard Model predictions.
\end{abstract}

\vspace*{2.0cm}

\begin{center}
  Published in Phys.~Lett.~B767 (2017) 110
\end{center}

\vspace{\fill}

{\footnotesize 
\centerline{\copyright~CERN on behalf of the \lhcb collaboration, licence \href{http://creativecommons.org/licenses/by/4.0/}{CC-BY-4.0}.}}
\vspace*{2mm}

\end{titlepage}

%%%%%%%%%%%%%%%%%%%%%%%%%%%%%%%%
%%%%%  EOD OF TITLE PAGE  %%%%%%
%%%%%%%%%%%%%%%%%%%%%%%%%%%%%%%%

%  empty page follows the title page ----
\newpage
\setcounter{page}{2}
\mbox{~}
%\newpage
%
%% Author List ----------------------------
%%  You need to get a new author list!
%% \input{LHCb_authors.tex}
%
%The author list for journal publications is provided by the Membership Committee shortly after 'approval to go to paper' has been given.
%%It will be made available on the page
%%\verb!http://www.physik.uzh.ch/~strauman/forMemCo/LHCb-PAPER-XXXX-XXX/! .
%It will be sent to you by email shortly after a paper number has beens assigned.
%The author list should be included already at first circulation, 
%to allow new members of the collaboration to verify whether they have been included correctly.
%Occasionally a misspelled name is corrected or associated institutions become full members.
%In that case, a new author list will be sent to you.
%In case line numbering doesn't work well after including the authorlist, try moving the \verb!\bigskip! after the last author to a separate line.
%
%
%The authorship for Conference Reports should be ``The LHCb
%  collaboration'', with a footnote giving the name(s) of the contact
%  author(s), but without the full list of collaboration names.

\cleardoublepage

%\twocolumn
% %%%%%%%%%%%%% ---------

\renewcommand{\thefootnote}{\arabic{footnote}}
\setcounter{footnote}{0}

%%%%%%%%%%%%%%%%%%%%%%%%%%%%%%%%
%%%%%  Table of Content   %%%%%%
%%%%%%%%%%%%%%%%%%%%%%%%%%%%%%%%
%%%% Uncomment next 2 lines if desired
%\tableofcontents
%\cleardoublepage

%%%%%%%%%%%%%%%%%%%%%%%%%
%%%%% Main text %%%%%%%%%
%%%%%%%%%%%%%%%%%%%%%%%%%

\pagestyle{plain} % restore page numbers for the main text
\setcounter{page}{1}
\pagenumbering{arabic}

%% Uncomment during review phase. 
%% Comment before a final submission.
%\linenumbers

\section{Introduction}
\label{sec:introduction}

The production of \ttbar pairs from proton-proton ($pp$) collisions in the forward region is of considerable interest, as it may be sensitive to physics beyond the Standard Model (SM)~\cite{th_tt1}. Furthermore, forward \ttbar events can be used to constrain the gluon parton distribution function (PDF) at large momentum fraction~\cite{th_tt3}. The \ttbar cross-section has been measured at ATLAS and CMS using several final states and at various centre-of-mass energies~\cite{atlas_tt_7_8_tev,cms_tt_8tev_ljets_new,cms_tt_8tev_new}. LHCb has also measured top quark production in the forward region in the $W\!+\!b$ final state~\cite{LHCb-PAPER-2015-022}.

Measurements of the production cross-sections of \wpb and \wpc in the forward region provide experimental tests of perturbative quantum chromodynamics (pQCD)~\cite{th_Wbb1,th_Wbb2,th_Wbb3}, in a complementary phase space region to ATLAS and CMS. Previous studies of the \wpb final state have been performed by ATLAS \cite{atlas_Wbb_7tev} and CMS \cite{cms_Wbb_7tev,cms_Wbb_8tev} at centre-of-mass energies $\sqs=7\tev$ and $8\tev$. LHCb has previously performed measurements of the production cross-sections of a \PW boson with at least one observed $b$ or $c$ jet~\cite{LHCb-PAPER-2015-021} at $7$ and $8 \tev$, and a \PZ boson with at least one $b$ jet at $7 \tev$~\cite{LHCb-PAPER-2014-055}. 

This Letter reports a study of events containing one isolated lepton (muon or electron) and two heavy-flavour tagged jets to measure the production cross-sections of \ttbar, \wppb, \wmpb, \wppc and \wmpc. The study of \wpc is the first of its kind.
Measurements are performed using a data sample corresponding to an integrated luminosity of $1.98\pm0.02\invfb$ of $pp$ collisions recorded at $8\tev$ during 2012 by the LHCb experiment.

\section{The LHCb detector and samples}
\label{sec:detector}
The \lhcb detector~\cite{Alves:2008zz,LHCb-DP-2014-002} is a single-arm forward spectrometer fully instrumented in the \mbox{pseudorapidity} range $2<\eta <5$, designed for the study of particles containing \bquark or \cquark quarks. The detector includes a high-precision tracking system consisting of a silicon-strip vertex detector surrounding the $pp$ interaction region, a silicon-strip detector located upstream of a dipole magnet with a bending power of about $4{\mathrm{\,Tm}}$, and three stations of silicon-strip detectors and straw drift tubes placed downstream of the magnet. The tracking system provides a measurement of momentum, \ptot, of charged particles with a relative uncertainty that varies from 0.5\% at low momentum to 1.0\% at 200\gev.\footnote{In this Letter natural units where $c$ = 1 are used.} The minimum distance of a track to a primary vertex (PV), the impact parameter (IP), is measured with a resolution of $(15+29/\pt)\mum$, where \pt is the component of the momentum transverse to the beam, in\,\gev.
Different types of charged hadrons are distinguished using information from two ring-imaging Cherenkov detectors. Photons, electrons and hadrons are identified by a calorimeter system consisting of scintillating-pad (SPD) and preshower (PRS) detectors, an electromagnetic calorimeter and a hadronic calorimeter. Muons are identified by a system composed of alternating layers of iron and multiwire proportional chambers. The online event selection is performed by a trigger, which consists of a hardware stage, based on information from the calorimeter and muon systems, followed by a software stage, which applies a full event reconstruction.
 
The $\W\to\Pmu\nu$ candidates are required to satisfy the hardware trigger requirement for muons, of having hits in the muon system corresponding to a high transverse momentum particle, and to satisfy the software trigger requirement of $\pt(\mu)>10\gev$. The $\W\to\Pe\nu$ candidates are required to satisfy the hardware trigger requirement for electrons of having an electromagnetic cluster of high transverse energy associated with signals in the PRS and SPD detectors, and the software trigger, which selects events with an electron with $\pt(\Pe)>15\gev$. A global event cut (GEC) on the number of hits in the SPD is applied in order to prevent high-multiplicity events from dominating the processing time of the reconstruction code.

Simulated event samples of \wj, \zj, \ttbar, single-top and diboson (\wz,\zz) production are generated using \pythia8~\cite{Sjostrand:2007gs,*Sjostrand:2006za} with a specific \lhcb configuration~\cite{LHCb-PROC-2010-056}. Event samples of \wpb, \wpc, \zpb and \zpc production are generated with \alpgen~\cite{paper-alpgen},  which includes tree-level contributions with up to four additional emissions of final state partons with respect to the leading-order diagram. \pythia8 is used to perform the hadronisation for these samples. 
The cross-sections of the simulated processes are calculated at next-to-leading-order (NLO) including spin correlation effects with \mcfm~\cite{paper-MCFM} using the CT10 PDF set~\cite{paper-CT10}.
Decays of hadronic particles are described by \evtgen~\cite{Lange:2001uf}, in which final-state radiation is generated using \photos~\cite{Golonka:2005pn}.
The interaction of the generated particles with the detector, and its response, are implemented using the \geant toolkit~\cite{Allison:2006ve, *Agostinelli:2002hh} as described in Ref.~\cite{LHCb-PROC-2011-006}.
Since neither showering nor hadronisation are included in \mcfm, an overall correction is calculated to compare the measurements with the predicted cross-section at particle-level. This is done by generating \wpb, \wpc and \ttbar events with \pythia8 with the CT10 \PDF set~\cite{paper-CT10} where the same acceptance requirements are applied.
The particle-level lepton momentum used here is the momentum after final-state radiation as implemented in \pythia8.

\section{Event selection}
\label{sec:sel}

Events are selected by requiring the presence of either a high-\pt muon or electron and two heavy-flavour tagged jets. The same fiducial definition for lepton and jets used in previous studies~\cite{LHCb-PAPER-2015-022,LHCb-PAPER-2015-021,LHCb-PAPER-2016-011} is applied. The lepton must have $\ptl>20\gev$ and $2.0<\eta(\lepton)<\eta_{\rm{max}}(\lepton)$, where $\eta_{\rm{max}}(\lepton)$ is $4.50$ for a muon candidate, corresponding to the muon identification system acceptance, and is $4.25$ for an electron candidate, corresponding to the electromagnetic calorimeter acceptance. 
The jets are required to have $\ptj>12.5\gev$ and $2.2<\etaj<4.2$. Due to the limited sample size to validate the heavy-flavour tagging algorithm for higher \pt jets~\cite{LHCb-PAPER-2015-016}, only jets with $\ptj<100\gev$ are considered. The lepton is required to be isolated from both jets using $\Delta\text{\it R}(\lepton,\mathrm{j})>0.5$, where $\Delta\text{\it R}=\sqrt{\Delta\Peta^2+\Delta\Pphi^2}$ is the distance between them in \Peta-\Pphi space and \Pphi is the azimuthal angle. %This requirement serves to remove the jet formed including the lepton. 
This requirement serves to remove the background formed by leptons coming from the same parton as the jets. The jets are also required to have $\Delta\text{\it R}(\ja,\jb)>0.5$, where \ja (\jb) is the highest (second highest) \pt jet of the pair.
Events with $\ptnu<15\gev$, where $\ptnu$ is the transverse component of $(\vec{p}({\lepton}) + \vec{p}(\ja) + \vec{p}({\jb}))$, are removed to reduce the contamination from events not containing a \PW boson.
If more than one ($\lepton+\ja+\jb$) candidate is found in the event, the candidate with highest \ptnu is selected.

Jets are reconstructed using a particle flow algorithm~\cite{LHCb-PAPER-2013-058} and clustered using the \antikt algorithm~\cite{antikt} with distance parameter $R=0.5$ as implemented in the \fastjet software package~\cite{fastjet}. As in Ref.~\cite{LHCb-PAPER-2013-058}, the jet energy is corrected to the particle level, excluding neutrinos, and the same jet quality requirements are applied.  Jets are heavy-flavour tagged, \ie as originating from a $b$ or $c$ quark, by the presence of a secondary vertex (SV) with $\Delta\text{\it R}<0.5$ between the jet axis and the direction of flight of the heavy-flavour hadron candidate, defined by the vector from the PV to the SV position.

The SV-tagger algorithm, described in detail in Ref.~\cite{LHCb-PAPER-2015-016}, uses two boosted decision trees (BDTs)~\cite{Breiman,AdaBoost}: one that separates heavy-flavour from light-parton jets (\bdtbcl) and one that separates $b$ jets from $c$ jets (\bdtbc). Both jets used in the analysis are required to have $\bdtbcl>0.2$, which gives a heavy-flavour tagging efficiency of about $50\%$ ($20\%$) for $b$ ($c$) jets and a misidentification probability of about $0.1\%$ for light jets.

In order to suppress the \zj background, events with an additional oppositely charged high-\pt lepton that fulfills the lepton requirements described above are vetoed. Backgrounds from misidentified leptons or semileptonic decays of heavy-flavour hadrons are suppressed by two requirements applied to the lepton: \ipl must be less than $0.04\mm$ and $\ptl/\ptlj>0.8$, where \lj is defined as a reconstructed jet with relaxed quality criteria that contains the lepton.

\section{Backgrounds}
\label{sec:bkg}

In both the electron and muon channels, the background processes include \zpb and \zpc production with $\PZ\to\Pmu\Pmu$ or $\PZ\to\Pe\Pe$, where one of the final state leptons is not reconstructed. $\PZ(\to\Ptau\Ptau)+\bbbar$ production is also considered, where at least one \Ptau decays to an electron or a muon. A small contribution of $\PZ\to\tau\tau$ produced in association with one $b$ or $c$ jet is also included. Other processes of $\PZ$ production associated to jets are negligible. 
Background contributions from $\PW(\to\ell\neu)$+jets where the event does not contain two $b$ jets, and $\PW(\to\tau\neut)+\bbbar$ where \Ptau decays to an electron or muon are also included. Single-top, $\PW(\to\ell\neu)\PZ(\to\bbbar)$ and $\PZ(\to\ell\ell)\PZ(\to\bbbar)$ production are considered as background processes. The expected yields of the background processes described above are obtained from NLO cross-sections. Weight factors are applied to compensate for residual differences between data and simulation for GEC, trigger and heavy-flavour tagging efficiencies. Further details about these factors and their uncertainties are given in Sec.~\ref{sec:systematics}.
 
The QCD multi-jet background, which includes lepton misidentification and semileptonic decays of a beauty or charm hadron, is estimated by using events which fail the $\ptl/\ptlj>0.8$ requirement. 
The QCD multi-jet background normalisation is adjusted in order to describe the event yield at $\ipl>0.04\mm$, after subtracting the non-QCD backgrounds obtained from simulation.

\section{Signal yield determination}
\label{sec:fit}

\subsection{Overview}

The data sample is split into four subsamples, according to the flavour and charge of the lepton (\mupm and \epm). 
A simultaneous fit to the distributions of four variables is performed to determine the \ttbar, \wppb, \wmpb, \wppc, and \wmpc yields in each sample. 
The four variables used in the fit are the invariant mass of the two jets (\mjj), the response of a multivariate classifier trained to distinguish between \ttbar and \wpb events and the multivariate discriminant classifier for each jet, \ja \bdtbc and \jb \bdtbc, trained to discriminate between $b$ and $c$ jets.
The expected background components are obtained from simulation, with the exception of the QCD multijet background.
The fitted signal yields are converted into cross-sections using simulation and data-driven efficiencies and the measurement of the integrated luminosity~\cite{LHCb-PAPER-2014-047}. The systematic uncertainties are included as nuisance parameters in the fit and propagated to the final result.

\subsection{Fit variables}

While \wpb and \wpc processes can be disentangled using the \bdtbc variables for both jets, the separation between \ttbar and \wpb or \wpc is obtained by using the \mjj variable and a multivariate discriminant, \ugb, constructed such that
its response is minimally correlated with \mjj ~\cite{Rogozhnikov:2014zea}. The variables \mjj,  \ja $\bdtbc$ and \jb $\bdtbc$ are found to be uncorrelated.
The \ugb response is trained in simulation using 11 kinematic variables of the lepton and jets: \ptl, $\eta(\lepton)$, $\pt({\ja})$, $\pt({\jb})$, $m(\ja)$, $m(\jb)$, \ptjj, $\Delta\text{\it R}(\ja,\jb)$, $\Delta\text{\it R}({\rm jj},\ja)$, $\Delta\text{\it R}({\rm jj},\jb)$ and $\cos(\theta_{{\rm jj}}({\lepton}))$, where $\theta_{{\rm jj}}(\lepton)$ is the lepton scattering angle in the dijet rest frame and ${\rm jj}$ represents the dijet system.
The muon and electron decay channels are trained separately.
Figure \ref{fig:ugbmjj} shows the correlation between the \ugb and the \mjj variables. In the fit
all variables are treated as uncorrelated; the effect of the observed small correlations is taken into account
in the systematic uncertainties of the results.

\newenvironment{figurespace}[1]{%
\begin{list}{}{%
\setlength{\belowcaptionskip}{#1}%
}%
\item[]}{\end{list}}

\begin{figurespace}{-4pt}
\begin{figure}
\begin{center}
\includegraphics[width=0.5\textwidth]{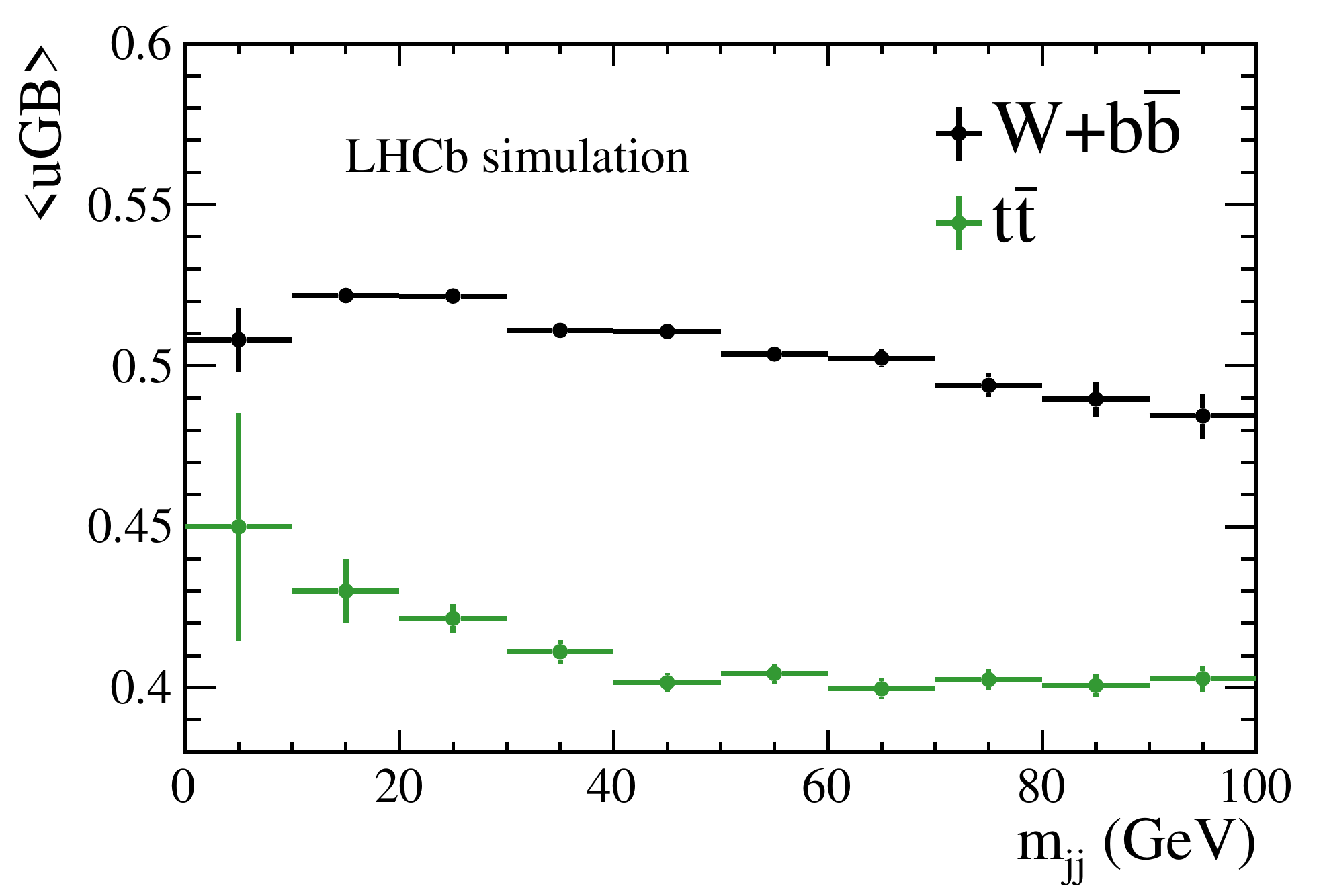}
\caption{Average of \ugb response in different intervals of \mjj for \wpb (black) and \ttbar (green). The vertical error bars represent the standard error of the \ugb mean in each interval.
\label{fig:ugbmjj}}
\end{center}
\end{figure}
\end{figurespace}

\subsection{Signal determination}

A binned maximum likelihood fit is performed to determine the yields of $\ttbar$, $\wppb$, $\wmpb$, $\wppc$ and $\wmpc$.  
The simulated background yields are normalised to NLO predictions and they are allowed to vary in the fit within their uncertainties. The QCD multijet background is normalised from a data-driven method as explained in Section~\ref{sec:bkg}. 
The fit is performed assuming the four variables (\mjj, \ugb, \ja $\bdtbc$ and \jb $\bdtbc$) to be uncorrelated. 

\begin{figure}
\begin{minipage}[b]{0.8\textwidth}
\includegraphics[width=\textwidth]{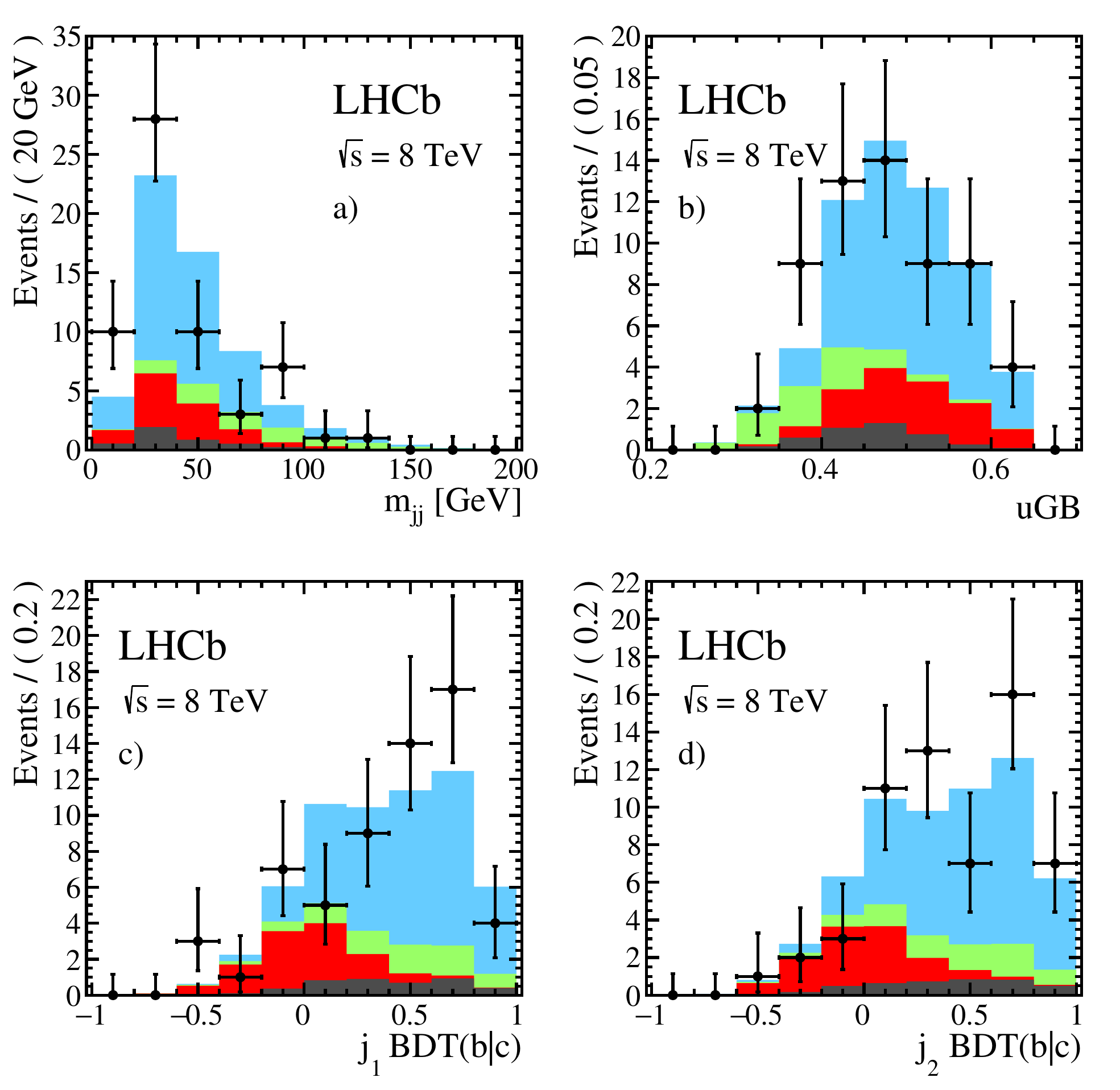}
\end{minipage}
\begin{minipage}[t]{0.19\textwidth}
\vspace{-8.cm}
\includegraphics[width=\textwidth]{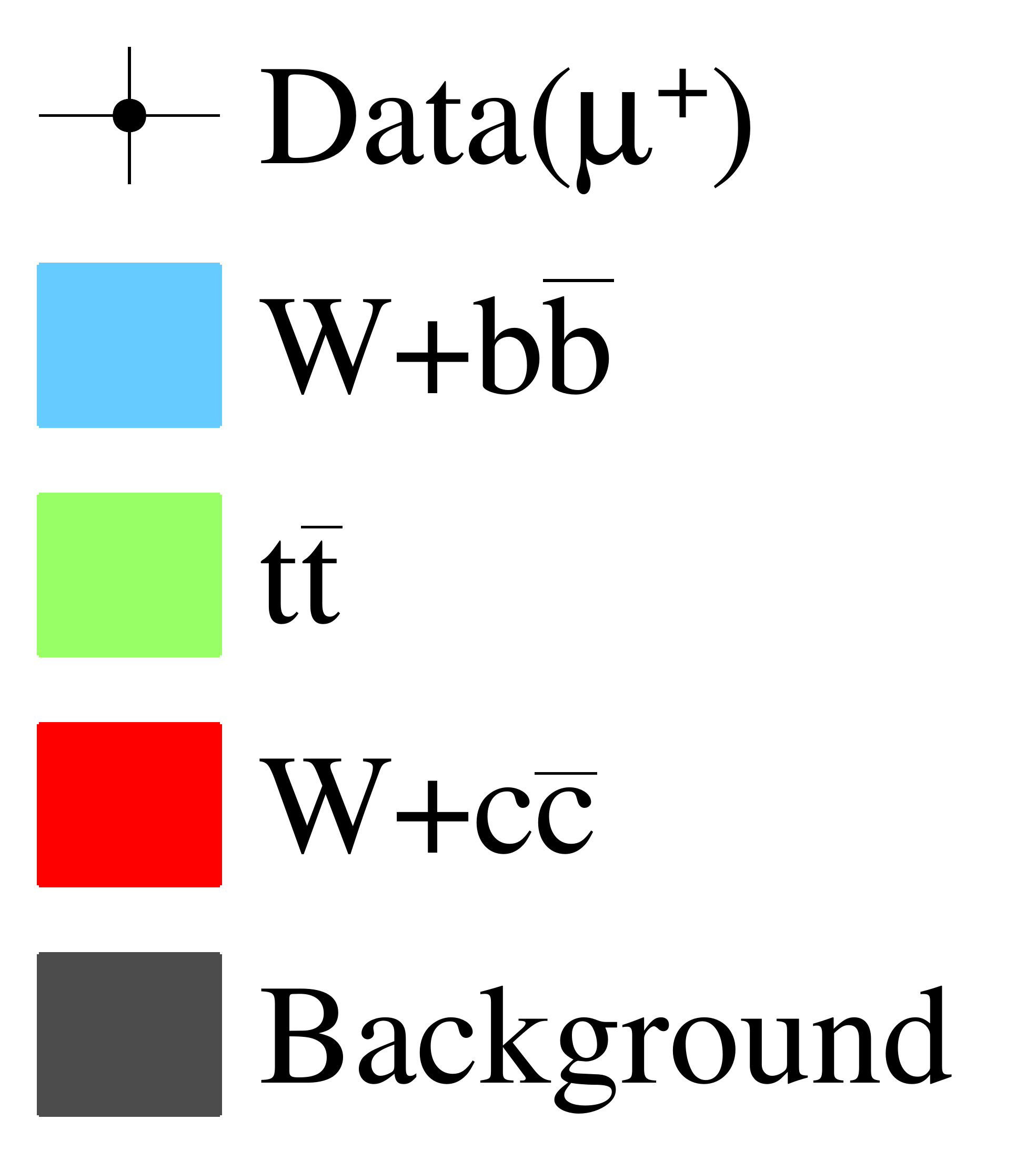}
\end{minipage}
\caption{Projections of the simultaneous 4D-fit results for the $\mu^+$ sample: a) the dijet mass; b) the \ugb response; the \bdtbc of the c) leading and d) sub-leading jets.  \label{fig:fit_muplus}}
\end{figure}

\begin{figure}
\begin{minipage}[b]{0.8\textwidth}
\includegraphics[width=\textwidth]{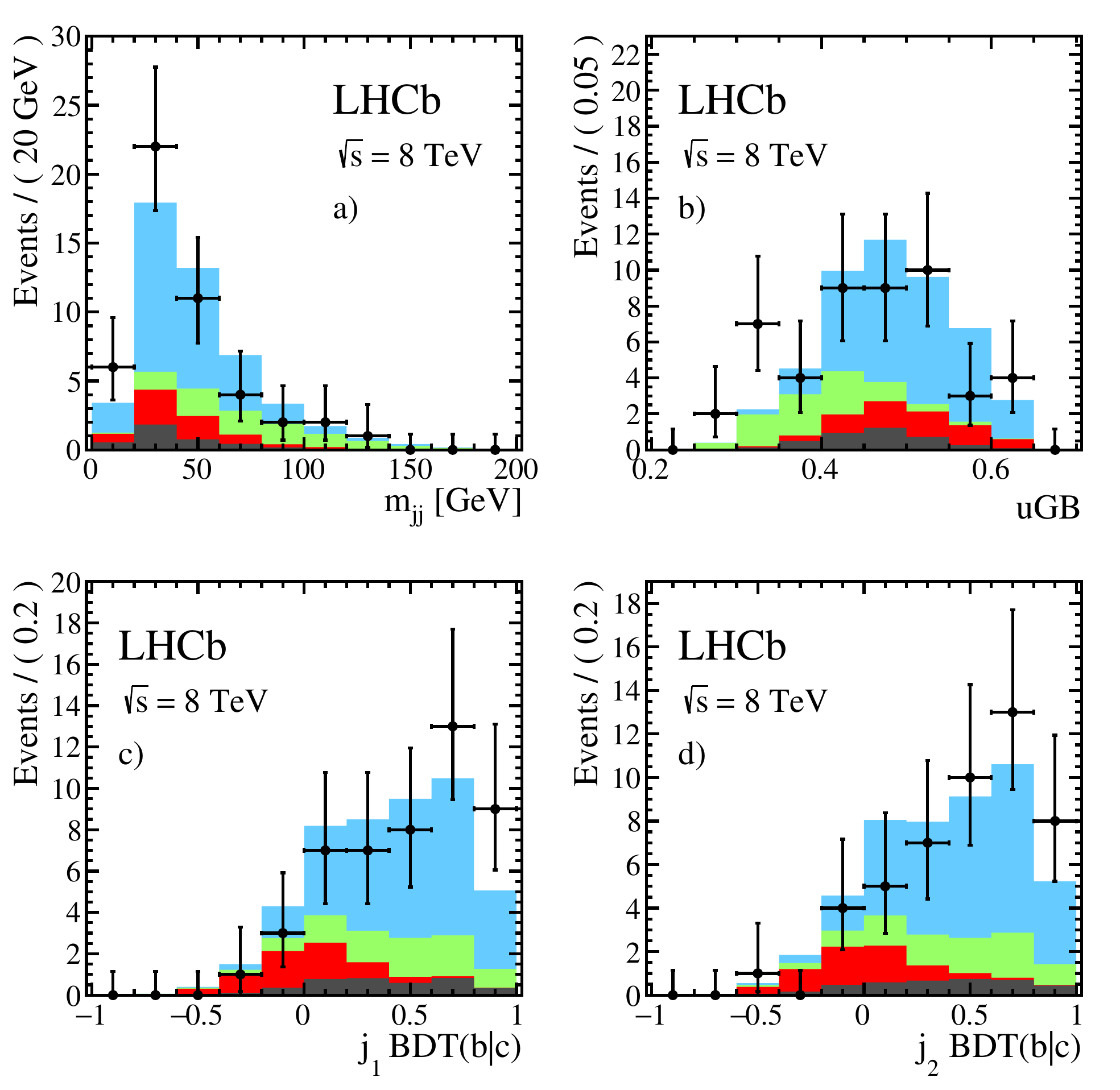}
\end{minipage}
\begin{minipage}[t]{0.19\textwidth}
\vspace{-8.cm}
\includegraphics[width=\textwidth]{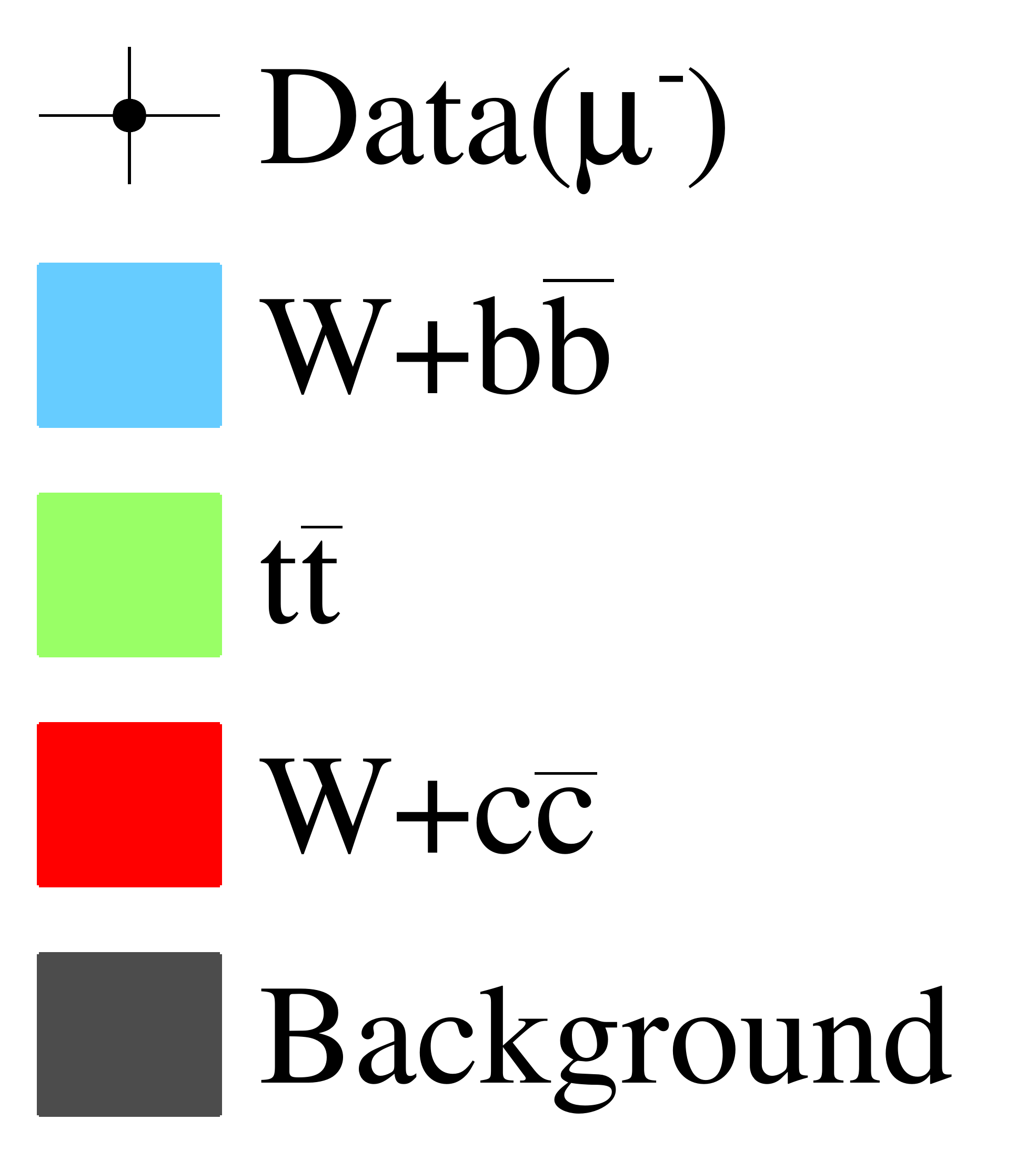}
\end{minipage}
\caption{Projections of the simultaneous 4D-fit results for the $\mu^-$ sample: a) the dijet mass; b) the \ugb response; the \bdtbc of the c) leading and d) sub-leading jets.  \label{fig:fit_muminus}}
\end{figure}

\begin{figure}
\begin{minipage}[b]{0.8\textwidth}
\includegraphics[width=\textwidth]{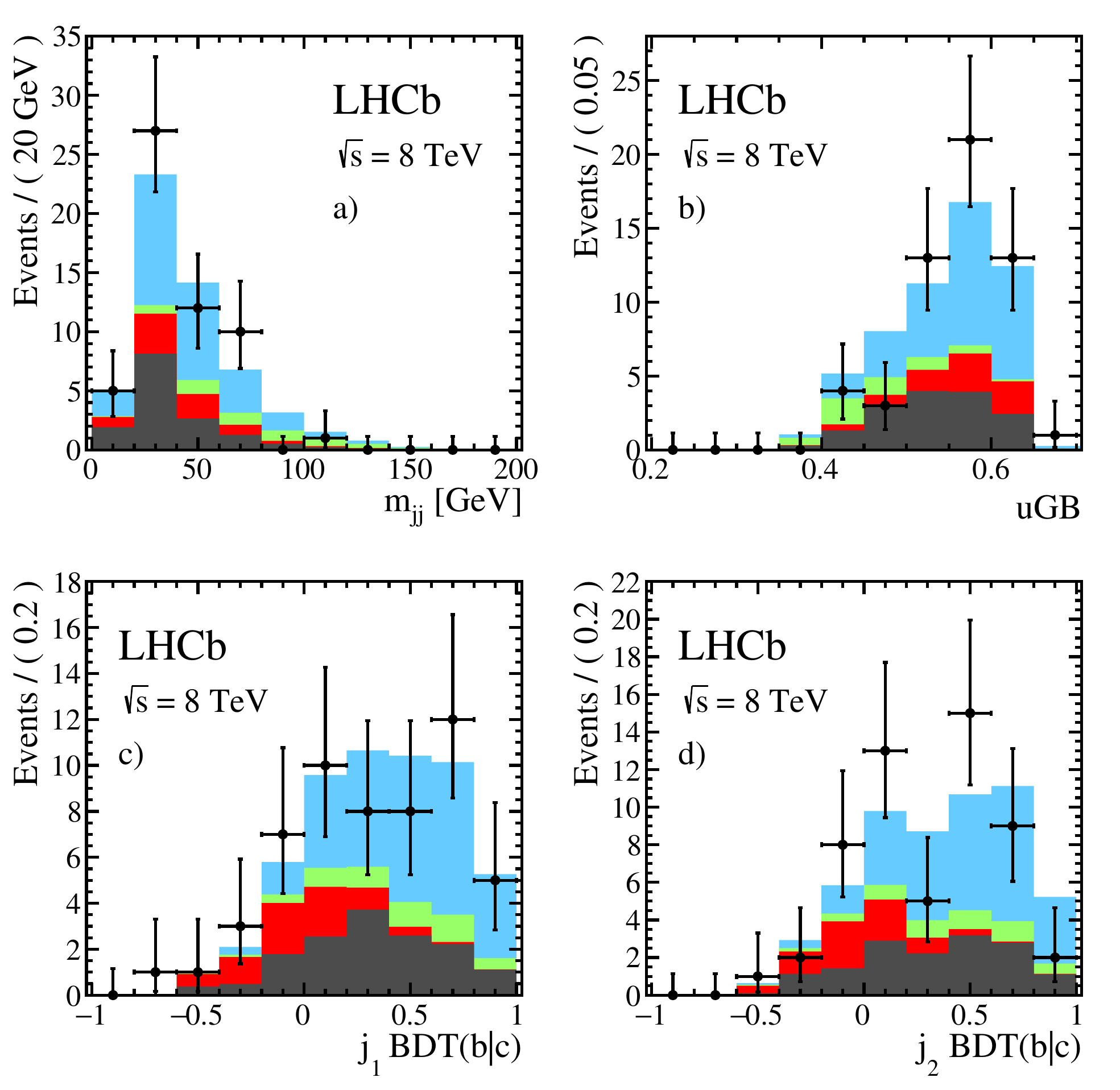}
\end{minipage}
\begin{minipage}[t]{0.19\textwidth}
\vspace{-8.cm}
\includegraphics[width=\textwidth]{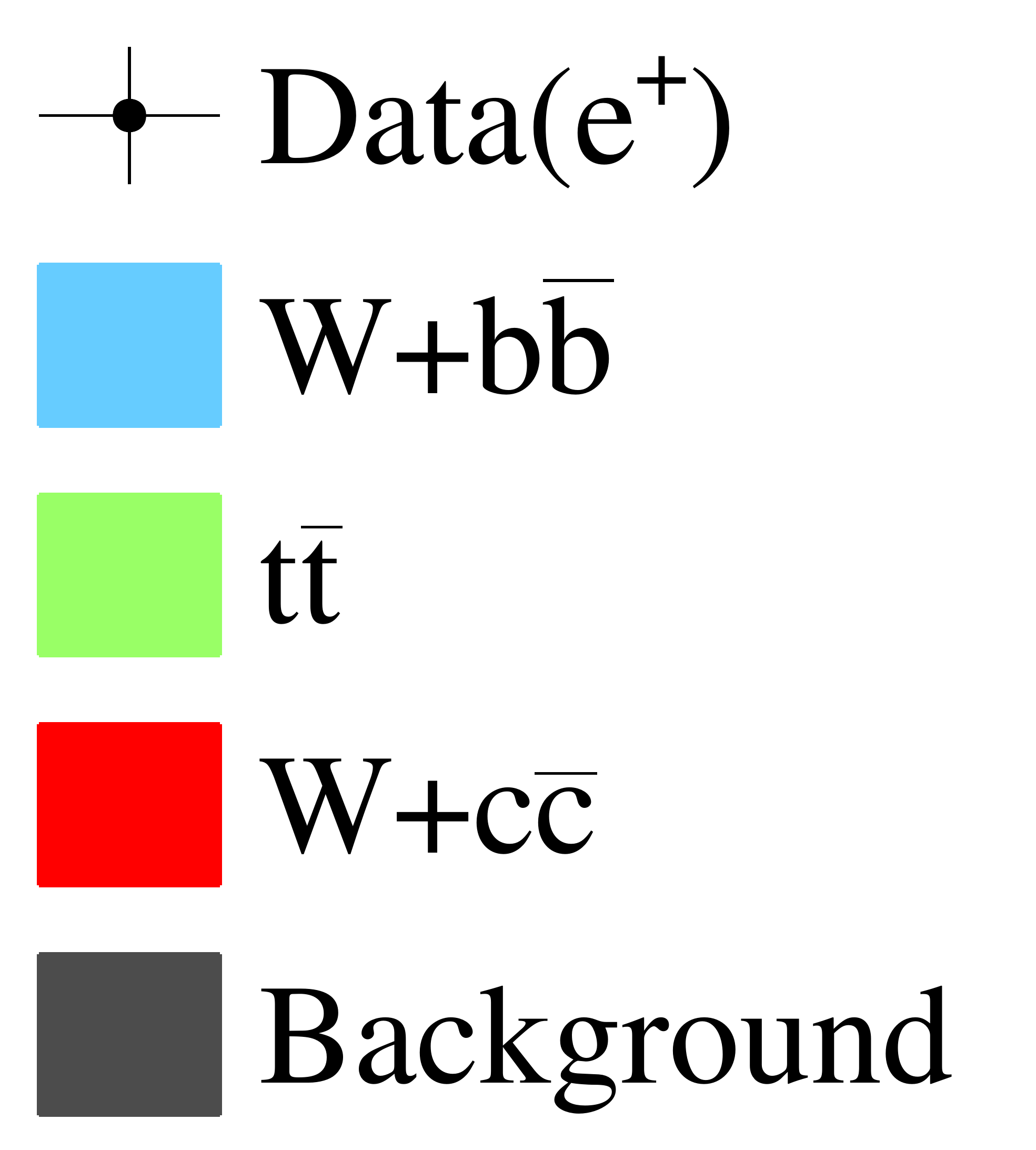}
\end{minipage}
\caption{Projections of the simultaneous 4D-fit results for the $e^+$ sample: a) the dijet mass; b) the \ugb response; the \bdtbc of the c) leading and d) sub-leading jets.  \label{fig:fit_elplus}}
\end{figure}

\begin{figure}
\begin{minipage}[b]{0.8\textwidth}
\includegraphics[width=\textwidth]{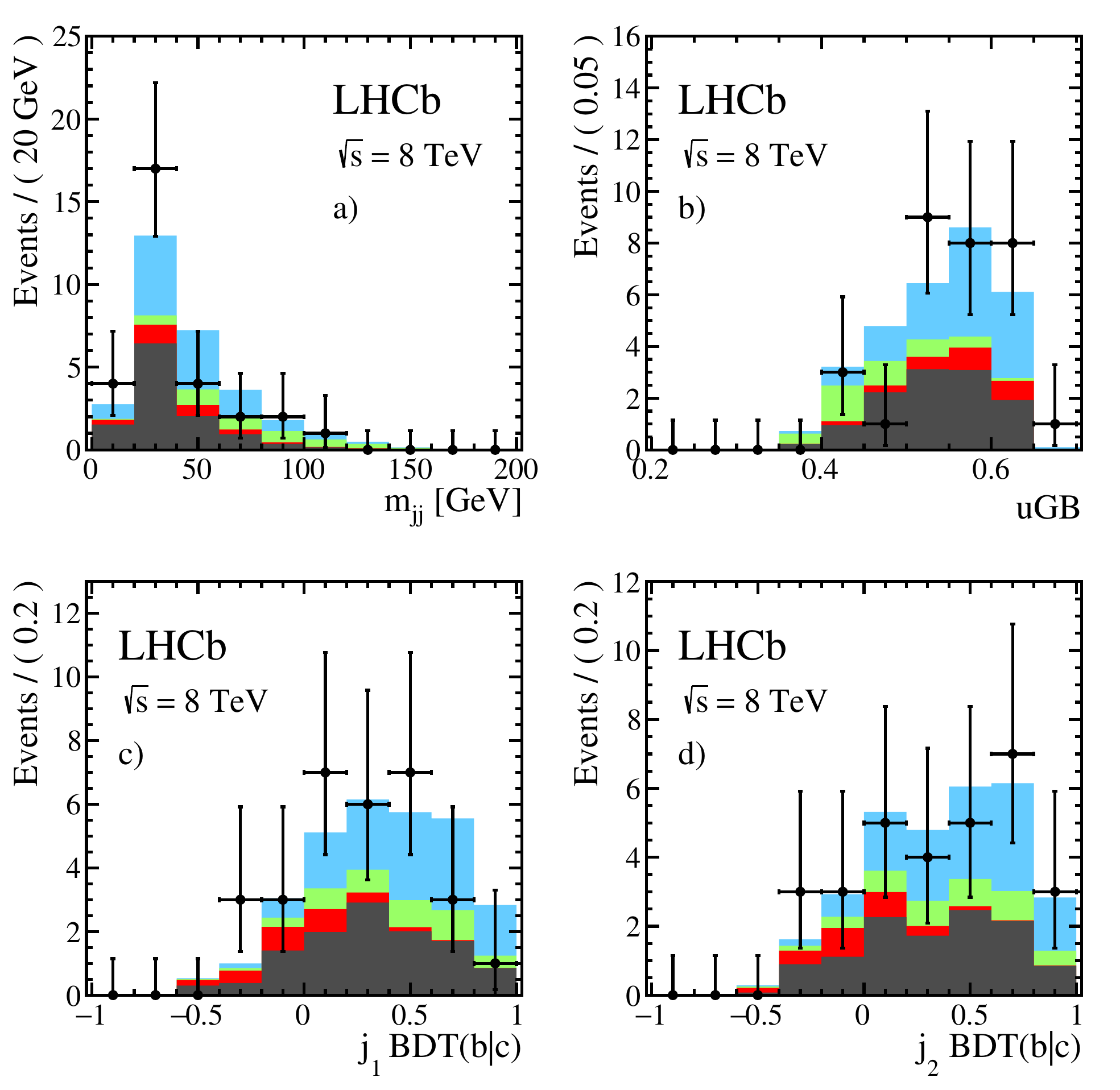}
\end{minipage}
\begin{minipage}[t]{0.19\textwidth}
\vspace{-8.cm}
\includegraphics[width=\textwidth]{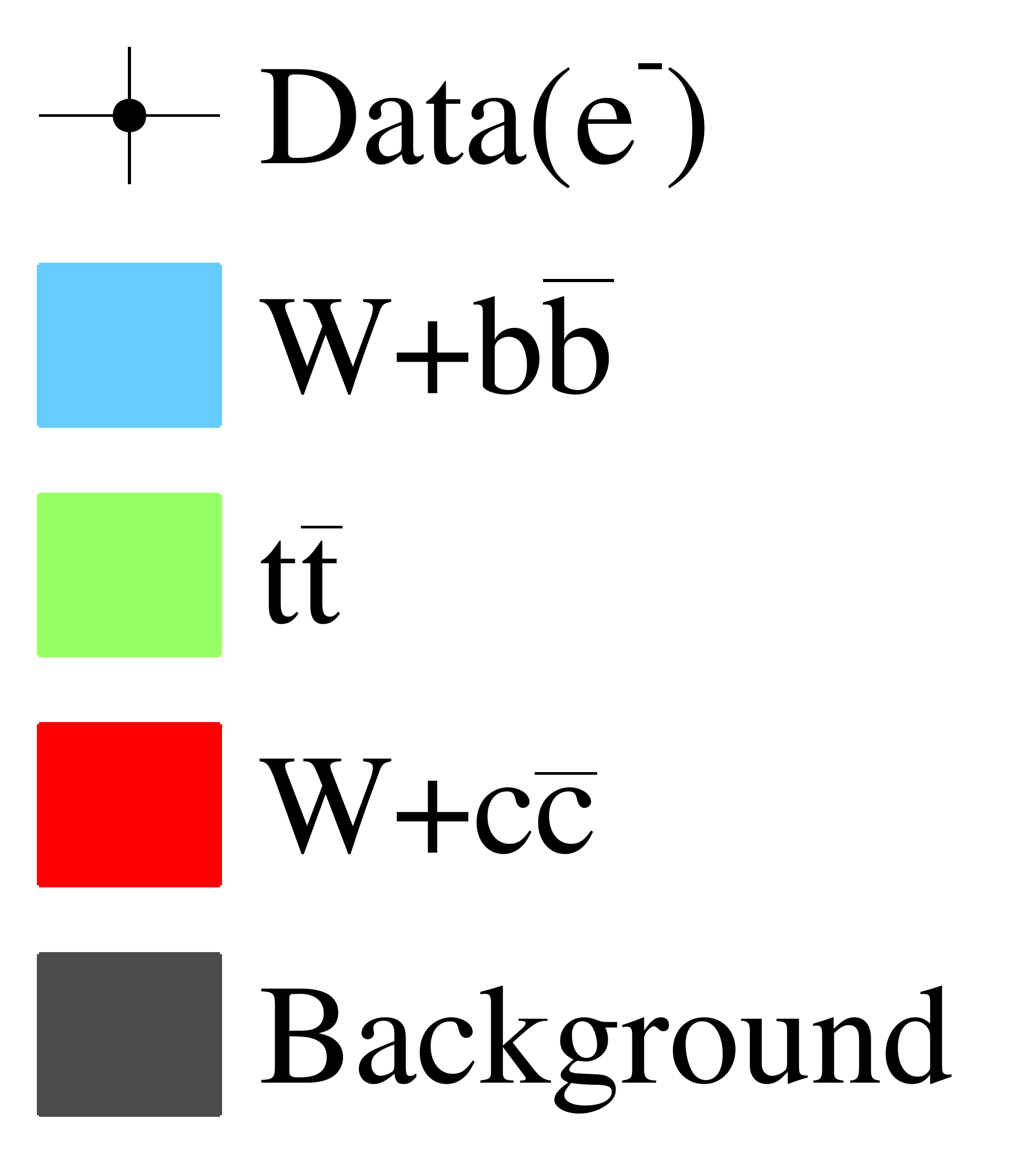}
\end{minipage}
\caption{Projections of the simultaneous 4D-fit results for the $e^-$ sample: a) the dijet mass; b) the \ugb response; the \bdtbc of the c) leading and d) sub-leading jets.  \label{fig:fit_elminus}}
\end{figure}

The free parameters in the fit are the normalisation factors with respect to the SM predicted yields $K(i)$, where $i=\ttbar,\wppb,\wmpb,\wppc,\wmpc$.
The $K(\ttbar)$ parameter is fitted using all four samples, while the others are fitted in each corresponding sample. 
The projections of the fit in each of the four samples are shown in Figs.~\ref{fig:fit_muplus}-\ref{fig:fit_elminus}, while the fit results are given in Table~\ref{tab:fit_results}. 

\begin{table}[h]
\caption{Simultaneous 4D-fit results for each of the four signal categories ($e$ and $\mu$, negative and positive). The normalisation factor $K$ and the fitted yields per sample are shown. The uncertainties quoted are statistical only.\label{tab:fit_results}}
 \centering
 \bgroup
 \def\arraystretch{1.3}
\begin{tabular}{c|c|c|c}
\toprule
Signal & $K$ & \Pmu sample yields & \Pe sample yields \\  
\midrule
\wppb & $1.49^{+0.23}_{-0.22}$ &  $45.5^{+6.9}_{-6.4}$ & $20.5^{+3.1}_{-2.9}$ \\
\wmpb & $1.67^{+0.33}_{-0.30}$ & $28.7^{+5.6}_{-4.9}$ &  $12.1^{+2.3}_{-2.1}$ \\
\wppc & $1.92^{+0.68}_{-0.58}$  & $12.8^{+4.5}_{-3.9}$ & $5.7^{+2.0}_{-1.7}$ \\
\wmpc & $1.58^{+0.87}_{-0.73}$  & $5.7^{+3.1}_{-2.6}$  & $2.5^{+1.4}_{-1.2}$ \\
\multirow{2}{*}{\ttbar} & \multirow{2}{*}{$1.17^{+0.35}_{-0.31}$} & $8.7^{+2.6}_{-2.3}$ (\mup)  & $3.7^{+1.1}_{-1.0}$ (\ep)  \\
			&																							& $8.3^{+2.5}_{-2.2}$ (\mun)  & $4.0^{+1.2}_{-1.1}$ (\en) \\
\bottomrule
\end{tabular}
\egroup
\end{table}

\subsection{Systematic uncertainties}
\label{sec:systematics}

Systematic effects can impact the results in two ways: by affecting signal and background yields, or by altering template shapes used in the fits. 
The efficiency of the GEC is measured in a \zsj sample selected with a looser trigger requirement~\cite{LHCb-PAPER-2016-011} and a $2\%$ uncertainty is assigned to account for the final-state dependence of the GEC efficiency observed in simulation.
The systematic uncertainty on the integrated luminosity is $1.16\%$~\cite{LHCb-PAPER-2014-047}.

The lepton  reconstruction and trigger efficiencies are studied using data-driven methods in $\PZ\to\ellell$~\cite{LHCb-PAPER-2015-049,LHCb-PAPER-2015-003}. Those studies show that data and simulation agree within 1.0--$5.0\%$ depending on $\eta(\ell)$ and $\pt(\ell)$, which is taken as systematic uncertainty.
The uncertainty of the lepton kinematic efficiency, which includes the effect of final-state radiation, is neglected.
The method described in Ref.~\cite{LHCb-PAPER-2015-016} is used to assess the systematic uncertainty due to the errors of the heavy-flavour tagging efficiency weight-factor described in Sec.~\ref{sec:bkg}, which amounts to 5--$10\%$ depending on \ptj.

The systematic uncertainty of the jet energy calibration includes possible biases due to flavour dependence ($2\%$), tracks not associated to a real particle ($1.2\%$), track momentum resolution ($1\%$) and residual differences between simulation and data due to pile-up and calorimeter response ($1\%$) as described in Refs.~\cite{LHCb-PAPER-2016-011,LHCb-PAPER-2013-058}. The jet energy resolution at LHCb is modelled in simulation to an accuracy of about $10\%$~\cite{LHCb-PAPER-2013-058,LHCb-PAPER-2015-021}. The uncertainties related to the jet reconstruction and quality selection efficiencies are found to be below $2\%$. The jet-related systematic uncertainties affect both the template shapes and the expected yields.

The simulated background normalisations are predicted at NLO and they are affected by uncertainties on the PDF (\deltapdf), on the strong coupling constant \alphas (\deltaalphas) and on the renormalisation and factorisation scales ($\delta_{\rm scale}$). 
The PDF uncertainty is evaluated following the procedure of Ref.~\cite{pdf4lhc}.  The influence of the uncertainty on the strong coupling constant is evaluated by calculating the cross-sections with %three 
PDF sets~\cite{paper-CT10} using values of $\alphas(M_Z)$: 0.117, 0.118 and 0.119.
The scale uncertainty is evaluated by calculating the cross-sections varying the renormalisation and factorisation scale by a factor of two.
The total uncertainty is taken as $\sqrt{(\deltapdfsq+\deltaalphassq)}+\delta_{\rm scale}$ as done in Ref.~\cite{LHCb-PAPER-2015-022} which translates to  relative uncertainties on the signal yields in the range 3--$10\%$.
These theoretical uncertainties are also considered in the signal yields in the experimental acceptance.% expected SM cross-sections that are compared to the measurement.

The systematic uncertainties in the normalisations due to the limited size of the simulated samples are between 1 and 7\%. The uncertainty on the normalisation of the QCD multi-jet background, taken from data, is found to have a
negligible effect.

Possible correlation effects between the fitted variables are studied by using templates generated randomly from the analysis templates with or without correlations found in simulation. It is found that the correlation and the fit procedure can affect the final yields by up to $10\%$. 

All significant systematic uncertainties are correlated between the four samples except for the uncertainty due to the finite size of the simulated samples, which affects each sample and process independently.

\section{Results and Conclusions}
\label{sec:results}

\begin{figure}[h!]
\begin{center}
\includegraphics[width=0.95\textwidth]{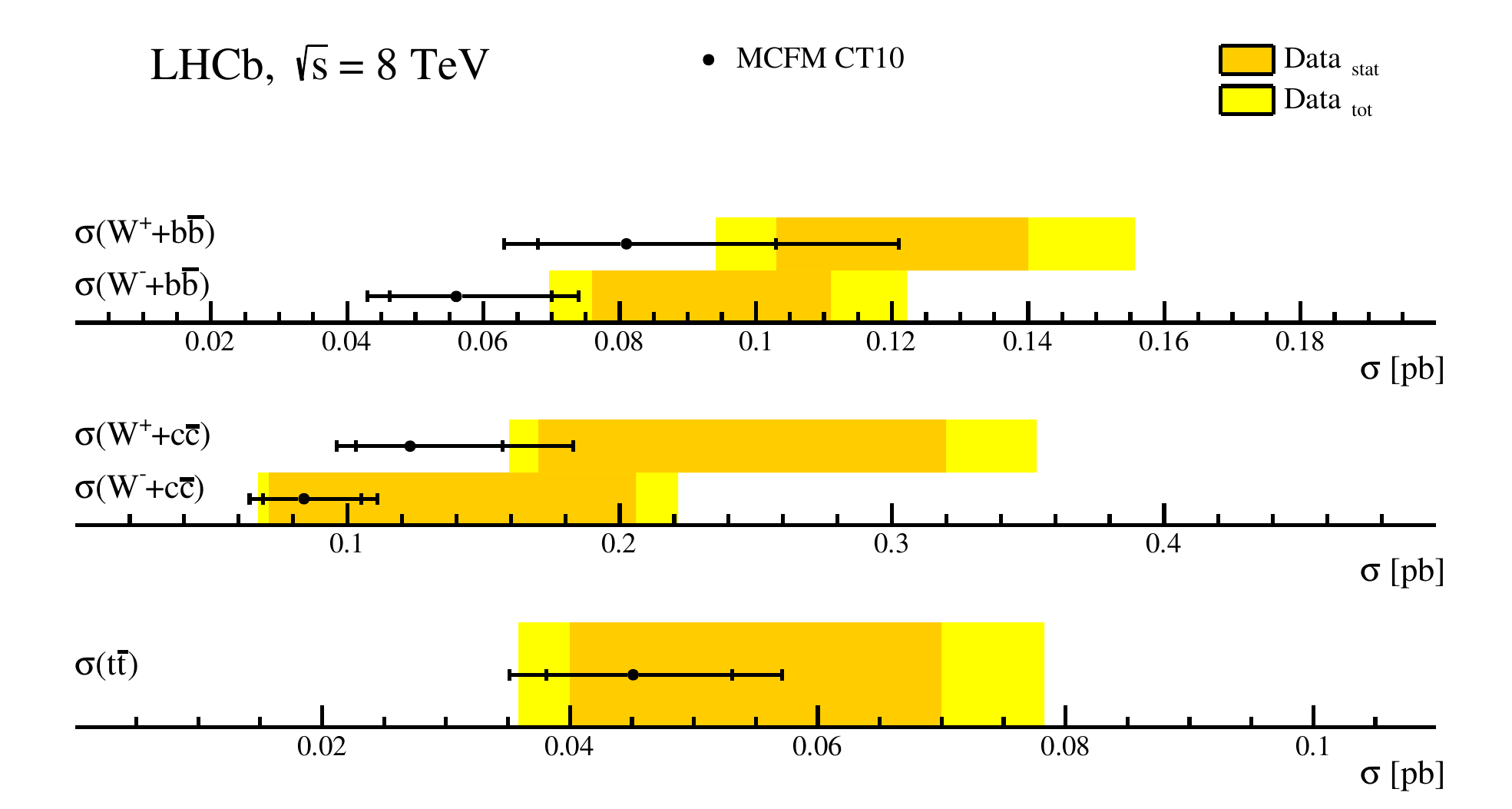}
\caption{Graphical representation of Table~\ref{tab:mxsec}. 
The outer bars (light yellow) correspond to the total uncertainties of the measured cross-sections and the inner bars (dark yellow) correspond to the statistical uncertainties. Theoretical prediction is represented by the black markers and error bars, where inner and outer uncertainties represent the scale and the total errors respectively.\label{fig:xs12} }
\end{center}
\end{figure}

The production cross-sections for \ttbar, $W^++\bbbar$, $W^-+\bbbar$, $W^++\ccbar$ and $W^-+\ccbar$ are measured for $pp$ collisions at a centre-of-mass energy of 8~\tev corresponding to an integrated luminosity of $1.98 \pm0.02 \invfb$ of data collected in 2012 by the LHCb experiment.
These production cross-sections are obtained as the product of the normalisation factors shown in Table~\ref{tab:fit_results} and the expected SM cross-sections.
The muons (electrons) coming from the \PW boson are required to have $2.0<\eta(\lepton)<4.5$ ($2.0<\eta(\lepton)<4.25$) and  $\ptl>20\gev$, while the jets are required to have $2.2<\etaj<4.2$ and  $\ptj>12.5\gev$. In addition, the
transverse component of $(\vec{p}({\lepton}) + \vec{p}(\ja) + \vec{p}({\jb}))$ is required to be $\ptnu>15\gev$.
The measured and expected cross-sections are presented in Table~\ref{tab:mxsec} and Fig.~\ref{fig:xs12}.
The significance obtained using Wilks' theorem \cite{Wilks:1938dza} is 4.9$\sigma$ for \ttbar, 7.1$\sigma$ for \wppb, 5.6$\sigma$ for \wmpb, 4.7$\sigma$ for \wppc and 2.5$\sigma$ for \wmpc. The correlation matrix of the measured cross-sections is presented in Table~\ref{tab:correlations}.
The measured cross-sections are in agreement with the SM predictions calculated at NLO using MCFM and the CT10 PDF set.

\begin{table}[h!]
  \begin{center}
	  \caption{\label{tab:mxsec} Observed and expected cross-sections in the fiducial region defined in Section~\ref{sec:sel}.
The first uncertainty on the expected cross-sections is related to the scale  variation and the second is the total. 
The first uncertainty on the observed cross-sections is statistical and the second is systematic.}
      \bgroup
      \def\arraystretch{1.3}
      \setlength\tabcolsep{1.5pt} 
      \begin{tabular}{c|lll|lll|c}
        \toprule
        Process ~ &  \multicolumn{3}{c|}{Expected $[\pb]$} &  \multicolumn{3}{c|}{Observed $[\pb]$} & ~ Significance ~\\

        \midrule
 \wppb ~ & ~ $0.081$&$^{+0.022}_{-0.013}$&$^{+0.040}_{-0.018}$ ~ & ~ $0.121$&$^{+0.019}_{-0.018}$&$^{+0.029}_{-0.020}$ ~ & $7.1 \sigma$\\
 \wmpb ~ & ~ $0.056$&$^{+0.014}_{-0.010}$&$^{+0.018}_{-0.013}$ ~ & ~ $0.093$&$^{+0.018}_{-0.017}$&$^{+0.023}_{-0.016}$ ~ & $5.6 \sigma$\\
 \wppc ~ & ~ $0.123$&$^{+0.034}_{-0.020}$&$^{+0.060}_{-0.027}$ ~ & ~ $0.24$&$^{+0.08}_{-0.07}$&$^{+0.08}_{-0.04}$ ~ & $4.7 \sigma$ \\
 \wmpc ~ & ~ $0.084$&$^{+0.021}_{-0.015}$&$^{+0.027}_{-0.020}$ ~ & ~ $0.133$&$^{+0.073}_{-0.062}$&$^{+0.050}_{-0.022}$ & $2.5 \sigma$\\
 \ttbar ~ & ~ $0.045$&$^{+0.008}_{-0.007}$&$^{+0.012}_{-0.010}$ ~ & ~ $0.05$&$^{+0.02}_{-0.01}$&$^{+0.02}_{-0.01}$ ~ & $4.9 \sigma$\\
         \bottomrule
       \end{tabular}
       \egroup
          \end{center}
\end{table}

\begin{table}[h!]
  \begin{center}
\caption{ Correlation matrix for the measured cross sections. The correlations are given in \%. \label{tab:correlations}}

      \bgroup
      \def\arraystretch{1.5}
      
\begin{tabular}{c|ccccc}
Process~&~ \ttbar ~& ~ \wppb ~&~\wmpb ~&~\wppc ~& ~\wmpc ~   \\ \hline
\ttbar ~& 100.00 &  &  &  & \\ 
\wppb ~& 39.02 & 100.00 &  &   &  \\ 
\wmpb ~& 35.10 & 58.62 & 100.00 &   &  \\ 
\wppc ~& 31.26 & 30.87 & 37.65 & 100.00 &  \\ 
\wmpc  ~& 19.06 & 31.97 & 20.16 & 22.99 & 100.00\\ 
\end{tabular}
      \egroup
\end{center}
\end{table}

\FloatBarrier

\section*{Acknowledgements}

\noindent We express our gratitude to our colleagues in the CERN
accelerator departments for the excellent performance of the LHC. We
thank the technical and administrative staff at the LHCb
institutes. We acknowledge support from CERN and from the national
agencies: CAPES, CNPq, FAPERJ and FINEP (Brazil); NSFC (China);
CNRS/IN2P3 (France); BMBF, DFG and MPG (Germany); INFN (Italy); 
FOM and NWO (The Netherlands); MNiSW and NCN (Poland); MEN/IFA (Romania); 
MinES and FASO (Russia); MinECo (Spain); SNSF and SER (Switzerland); 
NASU (Ukraine); STFC (United Kingdom); NSF (USA).
We acknowledge the computing resources that are provided by CERN, IN2P3 (France), KIT and DESY (Germany), INFN (Italy), SURF (The Netherlands), PIC (Spain), GridPP (United Kingdom), RRCKI and Yandex LLC (Russia), CSCS (Switzerland), IFIN-HH (Romania), CBPF (Brazil), PL-GRID (Poland) and OSC (USA). We are indebted to the communities behind the multiple open 
source software packages on which we depend.
Individual groups or members have received support from AvH Foundation (Germany),
EPLANET, Marie Sk\l{}odowska-Curie Actions and ERC (European Union), 
Conseil G\'{e}n\'{e}ral de Haute-Savoie, Labex ENIGMASS and OCEVU, 
R\'{e}gion Auvergne (France), RFBR and Yandex LLC (Russia), GVA, XuntaGal and GENCAT (Spain), Herchel Smith Fund, The Royal Society, Royal Commission for the Exhibition of 1851 and the Leverhulme Trust (United Kingdom).

\addcontentsline{toc}{section}{References}
\setboolean{inbibliography}{true}
\bibliographystyle{LHCb}
\bibliography{main,LHCb-PAPER}

\newpage

% Author List ----------------------------                                                                                                                                                                                                                                                                                                
%  You need to get a new author list!                                                                                                                                                                                                                                                                                                    

\centerline{\large\bf LHCb collaboration}
\begin{flushleft}
\small
R.~Aaij$^{40}$,
B.~Adeva$^{39}$,
M.~Adinolfi$^{48}$,
Z.~Ajaltouni$^{5}$,
S.~Akar$^{6}$,
J.~Albrecht$^{10}$,
F.~Alessio$^{40}$,
M.~Alexander$^{53}$,
S.~Ali$^{43}$,
G.~Alkhazov$^{31}$
% ...

{\footnotesize \it
% ...
$ ^{5}$Clermont Universit{\'e}, Universit{\'e} Blaise Pascal, CNRS/IN2P3, LPC, Clermont-Ferrand, France\\
$ ^{6}$CPPM, Aix-Marseille Universit{\'e}, CNRS/IN2P3, Marseille, France\\
% ...
$ ^{10}$Fakult{\"a}t Physik, Technische Universit{\"a}t Dortmund, Dortmund, Germany\\
% ...
$ ^{31}$Petersburg Nuclear Physics Institute (PNPI), Gatchina, Russia\\
% ...
$ ^{39}$Universidad de Santiago de Compostela, Santiago de Compostela, Spain\\
$ ^{40}$European Organization for Nuclear Research (CERN), Geneva, Switzerland\\
% ...
$ ^{43}$Nikhef National Institute for Subatomic Physics, Amsterdam, The Netherlands\\
% ...
$ ^{48}$H.H. Wills Physics Laboratory, University of Bristol, Bristol, United Kingdom\\
% ...
$ ^{53}$School of Physics and Astronomy, University of Glasgow, Glasgow, United Kingdom\\
% ...
}
\end{flushleft}

\end{document}